\documentclass[a4paper]{toledoStyle}
\usepackage{epsfig}

\newcommand{\Hd}{\ensuremath{\mathrm H_2}}

\newcommand{\CuoO}{\ensuremath{\mathrm C^{18} \mathrm O}}

\newcommand{\utCO}{\ensuremath{^{13}\mathrm{CO}}}

\def\NHt{NH$_3$}

\def\kms{~km~s$^{-1}$}
\def\cmmt{~cm$^{-3}$}
\def\cmmd{~cm$^{-2}$}
\def\smu{s$^{-1}$}              
\def\microns{$\mu$m}
\newcommand{\gsim}{\raisebox{-.4ex}{$\stackrel{>}{\scriptstyle \sim}$}}
  
\begin{document}

\title{The Galactic Center Interstellar Medium: from ISO to FIRST}

\author{N.J.\,Rodr\'{\i}guez-Fern\'andez \and J.M.\,Mart\'{\i}n-Pintado}

\institute{Observatorio Astron\'omico Nacional, Instituto Geogr\'afico 
Nacional, Apdo. 1143, E28800 Alcal\'a de Henares, Madrid, Spain} 

\maketitle 

\begin{abstract}
The {\it Infrared Space Observatory}\footnote{Based on observations with 
              ISO, an ESA project with instruments funded by ESA Member 
              States (especially the PI countries: France, Germany, 
              the Netherlands and the United Kingdom) and with the 
              participation of ISAS and NASA} 
(ISO) has shown the complexity of the Galactic center (GC) 
Interstellar medium (ISM)
detecting, not only large column densities of
 warm molecular gas (H$_2$),
but the emission of neutral atoms and ions of low ionization
potential (CII, OI, SiII,...)
that should arise in  shocked or  photon-dominated regions (PDRs).
In addition,
ISO has also detected emission from ions like SIII, NeII, ArII, or NII
(in some clouds we have even detected NeIII and OIII)
that  should arise from  HII regions that were previously unsuspected
due to the non-detection of Hydrogen  recombination lines.
Here we  review some  ISO results on the large
scale study of the GC ISM and in particular, on the heating mechanisms
of the clouds.                    
Although, shocks should play an important role on the physics and
chemistry of the GC ISM, ISO shows that the effect of  radiation
on the heating of the gas cannot be ruled out with the simple 
argument that the dust temperature is lower than that of the gas.
\end{abstract}
\section{Introduction}
In the few hundred central parsecs of the Galaxy (hereafter GC)
clouds are denser 
($n_\mathrm{H_2} \sim 10^4$ \cmmt ), more turbulent ($\Delta v \sim 20$ 
Km s$^{-1}$), and hotter than the clouds of the galactic disk.
There is a widespread warm gas component with temperatures of
100-200 K first known by NH$_3$ observations (see e.g.
\cite{rodriguezn:huttemeister93}) and now studied
in detail with  H$_2$ pure-rotational lines observations by ISO
(\cite{rodriguezn:RF01}, hereafter RF01).
The heating of the warm gas component over large regions where the
dust temperatures are much lower than those of the gas is a puzzle.
The discrepancy between the dust and the gas temperature is usually
considered to imply a mechanical, rather than radiative, heating mechanism
(\cite{rodriguezn:wilson82}, \cite{rodriguezn:gusten85}).
With the purpose of investigating the heating mechanisms of
the GC molecular clouds,
we have studied  a sample of 18  clouds distributed
all along the CMZ (Central Molecular Zone)
at millimeter and infrared wavelengths
using the ISO and the IRAM-30m telescopes.

ISO has allowed us to study the thermal balance of the GC clouds
by observing the major
coolants of the gas with temperatures of a few hundred Kelvin  like H$_2$, 
OI, or CII.
In the following we will review what we have learned from infrared (IR) 
and mm wavelength studies in the
field of the large scale study of the GC ISM.

\section{The molecular gas}
\label{sec:mol}
\subsection{Gas column densities from the CO data}
We have  observed the J=1-0 and J=2-1 lines of \CuoO~ and \utCO~ with
the IRAM-30m telescope.
The J=2-1 to J=1-0 line ratios are compatible with cold (20 K) and
dense gas (10$^4$  \cmmt) or warmer ($\sim$ 100 K) but less dense gas
(10$^3$ \cmmt).
However, the column densities derived in both cases are rather 
similar.
In general, if one considers a mixture of cold and warm gas
the total column densities traced by CO will be similar to those
derived for gas at 20 K,
which vary from source to source
but are in the range of $\sim 1-6~10^{22}$ \cmmd (RF01)

\subsection{Warm H$_2$}
We have observed several \Hd~  pure-rotational lines (from the S(0) to S(5)
lines) with the SWS spectrometer on board ISO.
The S(3) line is strongly absorbed by the 9.7 \microns~  
band of the silicates and has only been detected in the sources with
the most intense S(1) emission.
The visual extinction derived from the \Hd~  data is $\sim 30$ mag. 
After correcting for extinction one finds that the excitation temperature
derived from  the S(0) and S(1) lines ($T_{32}$) is between 130 and 200~K 
while that derived  from the S(4) and S(5) lines when detected is 500-700~K.
There is not a  clear dependence of $T_{32}$ on the distance 
to the Galactic center.
Extrapolating the populations in the J=2 and J=3 levels  (as derived from the
S(0) and S(1) lines) to the J=0 and J=1 levels at the temperature $T_{32}$ one
finds that the total warm \Hd~  column density varies from source to source but
is typically of 1-2 $10^{22}$ \cmmt.
The column density of gas at $\sim 600$ K is less than $1\%$ of the column 
density at $\sim 150$ K.
On average, the warm \Hd~  column densities are about $\sim 30 \%$ of the
total \Hd~ column densities derived from CO. 
For a few clouds the fraction of warm gas
is as high as 77$\%$ or even $\sim 100 \%$. This implies that for these
clouds almost all of the CO emission should arise from warm gas (see RF01).
Comparing with the \NHt~  observations of \cite*{rodriguezn:huttemeister93} one finds 
relatively high \NHt~  abundances of a few 10$^{-7}$ in both the warm and 
the cold gas.

\begin{figure}[ht]
\begin{center}
\epsfig{file=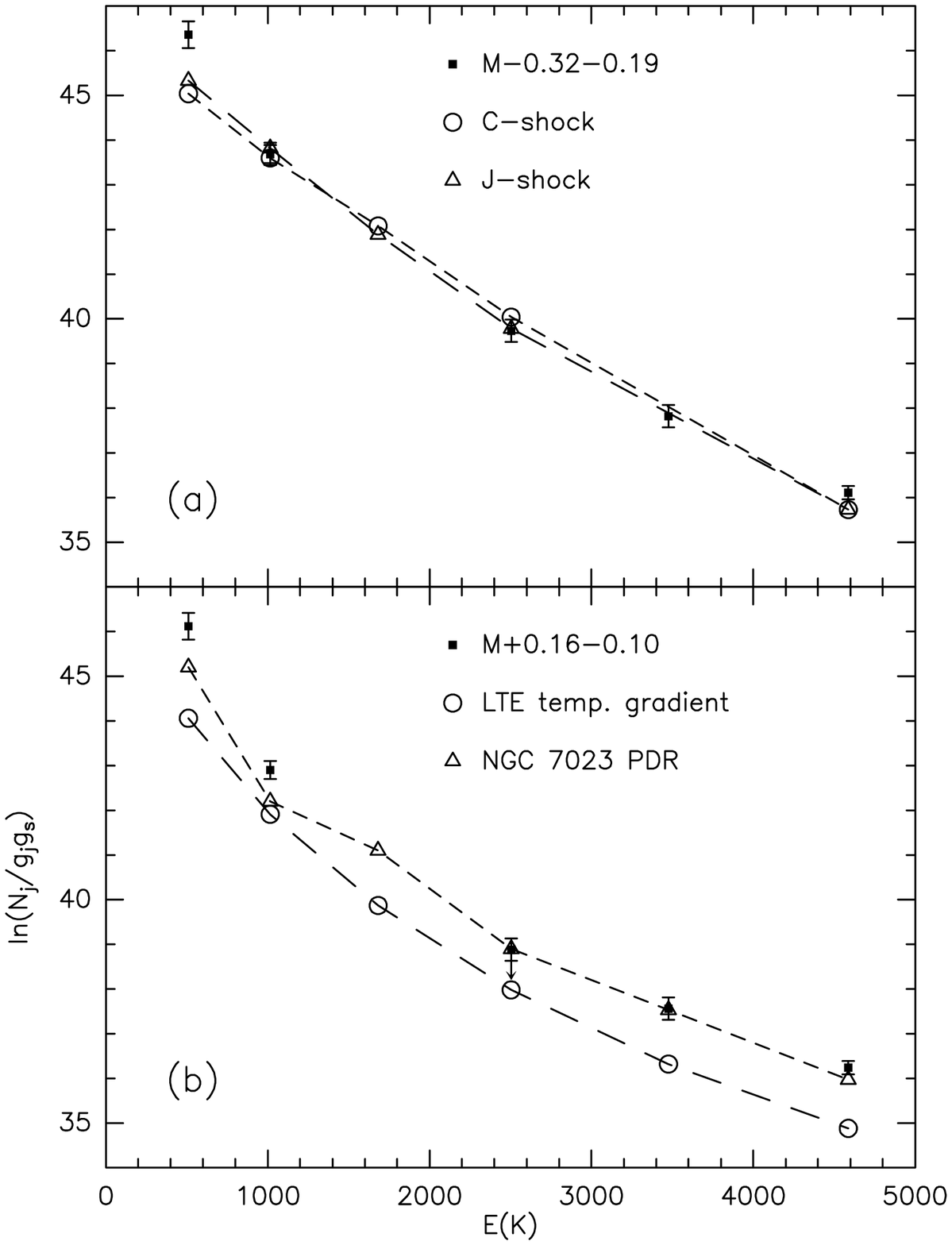, width=8.5cm}
\end{center}
\caption{ 
{\bf a)} Population diagram for M-0.32-0.19 (squares)
corrected for 30 mag. of visual extinction.
The errorbars represent upper limits to
the flux calibration uncertainties.
For comparison, it also displays the population diagrams
derived from the model of  Draine et al. (1983) of a shock
with velocity $\sim 12 $\kms and preshock density 10$^6$ \cmmt
(circles and dashed lines). Triangles and long-dashed lines
are used to plot the population diagram derived from the
J-shock model of Hollenbach \& McKee (1989) for a
velocity of 50 \kms and a preshock density of 10$^6$ \cmmt .
{\bf b)} Comparison of  the population diagram derived
for M+0.16-0.10 (squares) with the results of
Fuente et al. (1999) for the
NGC 7023 PDR (triangles and dashed line) and the population
diagram one obtains integrating the H$_2$ emission along
the temperature and H$_2$ abundance gradient derived by
Burton et al. (1990) for a PDR with density of 10$^6$ \cmmt\
and $G_0=10^4$ (open  circles)} 
\label{figexcith2}
\end{figure}

There are indirect arguments that points both to shocks  and 
to photo-dissociation regions (PDRs) as the heating mechanism
of the warm gas (see RF01 for a complete discussion).
Direct comparison of the \Hd~  data with PDRs and shocks models (see 
Fig.~\ref{figexcith2}) indicate that the S(4) and S(5) lines trace the 
densest gas in the GC clouds
(10$^6$ \cmmt) heated in PDRs, shocks, or both.
Nevertheless, to explain the large column densities of gas at $\sim 150$ K
traced by the S(0) and S(1) lines several less dense PDRs (with $G_0=10^3$ and
$n=10^3$ \cmmt) or low
velocity shocks ($< 10$ \kms) in the line of sight are required.
The curvature of the the population diagrams is in agreement with the
temperature gradient expected in a PDR (Fig.~\ref{figexcith2}b) but 
probably also with a composition of shocks with different velocities.
In summary, it is difficult to know if the \Hd~ emission arises in
PDRs or shocked regions.

\section{The ionized gas}
\label{sec:ion}
With both the LWS and SWS spectrometers on board ISO we have detected fine 
structure lines of ions like  [N II] 122~\microns, [SIII] 33 and 18~\microns~
or [Ne II] 12~\microns~ in most  sources. We have even detected
[N III] 57~\microns, [O III] 88 and 52~\microns~  or [Ne III] 15.6~\microns~
in the few sources located closer to the Galactic center. 
All these lines of ions with high excitational potentials
must arise in H II regions.
Otherwise, one needs to invoke J-shocks with velocities  \gsim 100 \kms~
and there is no hint of such strong shocks from the line profiles 
or the radial velocities of any observed line.
We derive effective temperatures for the ionizing radiation 
of $\sim 35000$~K, typical of an O7 star (see \cite{rodriguezn:MP99}, 2000).

We have  observed  the H35$\alpha$ and H$41\alpha$ recombination lines
in all the sources of our sample with the  IRAM-30m telescope. 
We have not detected any of these lines in any source (\cite{rodriguezn:MP99}, 2000).
Assuming line widths as large as 100 \kms~ and an electron temperature ($T_e$)
of 8000~K we can set a conservative {\it upper limit} to the number
of Lyman continuum photons  emitted per second by the excitation source of
$\sim 3~10^{47}$ \smu. This rate is typical of a star with a spectral type B0
or later and effective temperature $\le 30000$ K.

This implies that the number of Lyman continuum photons derived from
the radio recombination lines is not the total number of those
photons emitted by the ionizing  source.
The apparent inconsistency between the radio recombination lines and the fine
structure lines disappears if one considers that the HII regions are more
extended than the 30-m telescope beam, 
e.g., that the ionizing stars are surrounded by cavities with sizes of  $\sim
1-2$ pc (\cite{rodriguezn:MP99}, 2000).

\section{The dust}
\label{sec:dust}
Figure \ref{figdustall} shows the dust continuum spectra toward a 
representative sample of sources. 
Any other spectrum is very similar to one of those in 
Fig.~\ref{figdustall}.
The wavelength of the maximum of emission varies from $\sim 110$~\microns~
for M+0.76-0.05 and other sources with galactic longitude $> 1$\degr~
to $\sim 80$~\microns~ for the sources located in the Sgr C region and
even to $\sim 55$~\microns~ for a few sources located close to the Radio Arc.
\begin{figure}[ht]
\begin{center}
\vspace{5cm}
\end{center}
\caption{ LWS full grating spectra toward some of the clouds. A constant
has been added to some of them for display purposes}
\label{figdustall}
\end{figure}

To explain  the dust emission one needs at least two grey bodies.
Figure~\ref{figfitdust} displays the dust emission of 
the sources with coldest (M+0.76-0.05) and warmest dust (M+0.21-0.12)
together with the predictions of a model with two grey bodies at
different temperatures. 
For the predictions we have used the following expression:
\begin{equation}
S_\lambda=\Omega [f_c (1-e^{-\tau_c}) B(T_c,\lambda)+ f_w 
  e^{-\tau_c} (1-e^{-\tau_w}) B(T_w,\lambda)]
\end{equation}  
where $B(T,\lambda)$ represents the Planck function and the opacity,
$\tau$, is given by:
\begin{equation}
 \tau(\lambda)= 0.014 A_V  (30/\lambda)^\alpha 
\end{equation}
where $A_V$ is the visual extinction in magnitudes and $\lambda$ expressed
in \microns.  
This model assumes
that the warmer component is being extinguished by the colder one.
Taking $\Omega = 80^{''}\times 80^{''}$ (the LWS beam) we obtain
good fits for the values of the parameters listed  Table~\ref{tabfit}.
\begin{table}[bht]   
 \caption{ Parameters of the grey bodies shown in Fig.~\ref{figfitdust}}
 \label{tabfit}
 \begin{center}
 \leavevmode 
 \footnotesize
 \begin{tabular}[h]{lllll}
   \hline \\[-5pt]
   Source   &  $A_v$     &  $\alpha $    & $ T$ & $f$\\[+5pt]
   \hline \\[-5pt]
   M+0.76-0.05& 49     & 1.5 & 16 &10\\
              & 7.5    & 1   & 30 & 1 \\
   M+0.21-0.12& 50     & 1.5 & 16 & 6 \\           
              &2.2     & 1   & 47 & 1 \\
   \hline \\
 \end{tabular}
 \end{center}
\end{table}
To explain  the emission at large $\lambda$'s it  is necessary a grey body 
with temperature $\sim 15$ K. The visual extinction associated to
this component
would be 30--50 mag. To explain the large fluxes observed at
large $\lambda$'s,  filling factors ($f$) $>1$  are needed.
The temperature of the  cold component does not vary much from source
to source.

In addition to the cold dust component a warmer one is also  required.
This component would fill the beam and its temperature varies from
source to source from the $\sim 30$ K of M+0.75-0.05, M+3.06+0.34
and M+1.56-0.30 (\cite{rodriguezn:RF00}) to the $\sim 45$ K  of M+0.21-0.12 and
M+0.35-0.06. 
The visual extinction associated to  this component
would be $\sim$ 2-8 mag, that is, 5-10 $\%$ of the extinction due
to the cold dust component.

Clearly, a heating mechanism is needed to raise the dust temperature
from $\sim 15$ K to 30 K or even 45 K.
For the standard dust-to-gas ratio the extinction caused
by the warm dust component is equivalent to a \Hd~
column density of 2-8 $10^{21}$  \cmmd.
These equivalent column densities are only a factor of 2 lower than
the column densities of warm gas as derived from the \Hd~ pure-rotational 
lines.
The simplest explanation is        that
both the warm gas and dust arise in low-density PDRs like 
those modeled by \cite*{rodriguezn:hollenbach91}.
\begin{figure}[ht]
\begin{center}
\vspace{5cm}
\end{center}
\caption{LWS spectra (dots) and fits  to M+0.21-0.12 (blue) and 
 M+0.76-0.05 (black).
 Dotted and solid  lines represent a cold and a warm
 grey body, respectively, with parameters given in Table~\ref{tabfit}
 The curves
 plotted with empty squares are  the total emissions as defined by Eq.~1}
\label{figfitdust}
\end{figure}   

In the external layers of the PDR  the gas is heated
via photoelectric effect in the grain surfaces without heating the
dust to high temperatures.
For instance, in their ``standard" model ($n=10^3$ \cmmt~ and  $G_0=10^3$),
the gas can reach temperatures of 100--200 K in the first 3 mag
of visual extinction into the cloud where on average the dust 
temperature would be  $\sim 35$ K (see Fig. 3 of \cite{rodriguezn:hollenbach91}).
Those values for the PDR parameters are quite reasonable for the GC ISM
at large scale based on average densities and the far-IR continuum
(see discussion in \cite{rodriguezn:pak}).
PDR models with $n\sim10^3$ \cmmt~ and $G_0\sim10^3$ can also explain 
the \Hd~ excitation temperatures traced by the S(0) and S(1) lines
and the large scale emission of the \Hd~$v=1-0$ S(1) line (\cite{rodriguezn:pak}).

In this scenario, the warm dust component would be located at the GC while
the cold dust emission would arise both from  the GC clouds and from clouds in
the line of sight (typically $\sim 25$ mag of visual extinction are due
to material located between us and the GC region).

\section{Conclusions}                 
ISO has, for the first time, measured directly the column densities 
of warm molecular gas in the GC.
On average, the warm \Hd~ represents a fraction
of $\sim 30 \%$ of the total \Hd~ column densities derived from \utCO~ and 
\CuoO~ observations.

ISO has also detected fine structure lines of ions with high ionization 
potential that should arise from  HII regions.
The comparison of the fine structure lines with Hydrogen recombination
lines at mm wavelengths shows evidences 
for an extended ionized component with high effective radiation temperature.

This suggest the presence of a PDR in the interface 
between the HII region and the 
molecular material that should contribute to the heating of the
warm gas and the dust component with temperatures of 30-45 K.

There are many evidences that shocks play an important role in the 
physics and chemistry of the GC ISM.
The high abundances of molecules which are easily photodissociated
like \NHt~, SiO (\cite{rodriguezn:MP97}) and  C$_2$H$_5$OH (\cite{rodriguezn:MP01}) suggest that
they should be  sputtered from the grains by     shocks.
Nevertheless, ISO points out that the effect of UV radiation on the GC ISM
(and in particular on the heating of the molecular gas) cannot be ruled      
out with the simple argument that the dust temperature is lower than that
of the gas.

FIRST will    help to understand the energetics of the GC ISM. 
For instance,  the high
spectral and spatial resolution that the HIFI instrument will achieve,
will  be of great interest to study the interplay between the ionized
material and the warm molecular gas.
\begin{acknowledgements}
NJR-F acknowledges {\it Consejer\'{\i}a de Educaci\'on  Cultura
de la Comunidad de Madrid} for a pre-doctoral fellowship.
This work has been partially supported by
the Spanish CICYT and the European Commission
under grant numbers ESP-1291-E and 1FD1997-1442.
\end{acknowledgements}

\end{document}